\documentstyle[aps,prl,twocolumn]{revtex}
\input epsf
\def\c60{C$_{60}$}
\def\mgb2{MgB$_2$}
\def\2th{2$\theta$}
\begin{document}
\twocolumn[\hsize\textwidth\columnwidth\hsize\csname@twocolumnfalse\endcsname
\title{Compressibility of the \mgb2 Superconductor}
\author{K. Prassides,$^{1,2,3}$ Y. Iwasa,$^1$
T. Ito,$^1$ Dam H. Chi,$^1$ K. Uehara,$^1$ E. Nishibori,$^4$
M. Takata,$^4$ S. Sakata,$^4$ Y. Ohishi,$^5$ O. Shimomura,$^6$
T. Muranaka,$^7$ and J. Akimitsu$^{7,8}$}
\address{$^1$Japan Advanced Institute of Science and Technology,
Tatsunokuchi, Ishikawa 923-1292, Japan}
\address{$^2$School of Chemistry, Physics and Environmental Science,
University of Sussex, Brighton BN1 9QJ, UK}
\address{$^3$Institute of Materials Science, NCSR "Demokritos", 
153 10 Ag. Paraskevi, Athens, Greece}
\address{$^4$Department of Applied Physics, Nagoya University, Nagoya 
464-8603, Japan}
\address{$^5$Spring-8, Japan Synchrotron Radiation Research Institute, 
Hyogo 679-5198, Japan}
\address{$^6$SPring-8, Japan Atomic Energy Research Institute, 
Hyogo 679-5148, Japan}
\address{$^7$Department of Physics, Aoyama-Gakuin University, Chitosedai, 
Setagaya-ku, Tokyo 157-8572, Japan}
\address{$^8$CREST, Japan Science and Technology Corporation (JST), 
Tokyo, Japan}
\date{February 27, 2001}

\maketitle

\begin{abstract}

Considerable excitement has been caused recently by the discovery that 
the binary boride system with stoichiometry \mgb2 is superconducting 
at the remarkably high temperature of 39 K \cite{nature}. This 
potentially opens the way to even higher T$_c$ values in a 
new family of superconductors with unexpectedly simple composition 
and structure. The simplicity in the electronic and crystal structures 
could allow the understanding of the physics of high-T$_c$ 
superconductivity without the presence of the multitude of 
complicated features, associated with the cuprates. Synchrotron 
X-ray diffraction was used to measure the isothermal compressibility 
of \mgb2, revealing a stiff tightly-packed incompressible solid with 
only moderate bonding anisotropy between intra- and inter-layer directions. 
These results, combined with the pressure evolution of the 
superconducting transition temperature, T$_c$ establish its 
relation to the B and Mg bonding distances over a broad range of values. 
\\
\\
\end{abstract}

]


\mgb2 adopts a hexagonal crystal structure (AlB$_2$-type, space group 
$P$6/$mmm$) \cite{jacs} which is analogous to intercalated graphite with 
all hexagonal prismatic sites of the primitive graphitic structure 
(found in hexagonal BN) completely filled and resulting in two interleaved 
B and Mg layers. In addition, allowing for full charge transfer from 
Mg to the boron 2D sheets, the latter are themselves isoelectronic 
with graphite. 

Detailed information on the properties of \mgb2 is being currently 
rapidly accumulated. Band structure calculations clearly reveal that, 
while strong B-B covalent bonding is retained, Mg is ionized and 
its two electrons are fully donated to the B-derived conduction band 
\cite{theory1,theory2,theory3,theory4}. Superconductivity in \mgb2 
is then essentially due to the metallic nature of the boron 2D 
sheets and the presence of strong electron-phonon interactions 
together with the high vibrational frequencies of the light B atoms 
ensure a high transition temperature \cite{theory1}. Support for 
such a phonon-mediated BCS-type mechanism has been provided by 
measurements of the boron isotope effect ($\Delta$$T_c$= 1.0 K, 
isotope exponent $\alpha_B$$\sim$ 0.26) \cite{isotope}. 
In addition, $T_c$ has been found to decrease with applied pressure 
at the rate of -d$T_c$/d$P$$\sim$ 1.6 K/GPa up to 1.84 GPa \cite{highp}, 
again consistent with mediation of the pairing interaction by phonons. 
An alternative scenario derives from the fact that \mgb2 is hole-doped 
and superconductivity may be understood within a formalism developed for 
high-$T_c$ cuprate superconductivity \cite{hirsch}. Such a theory 
predicts a positive pressure coefficient on $T_c$ as a result 
of the decreasing intraplane B-B distance with increasing pressure 
and appears to disagree with the experimental observations. However, 
the response of the system may be more complex if pressure also 
affects the charge transfer between the B planes and Mg and will 
vary depending on whether the system is in the overdoped or underdoped regime.

Here, we address the problem of the evolution of the structural properties 
of the \mgb2 superconductor with applied pressure using synchrotron 
X-ray powder diffraction techniques. We find that \mgb2 remains 
strictly hexagonal to the highest pressure used. The isothermal 
interlayer compressibility,  dln$c$/d$P$ at zero pressure is only 
1.4 times the value of the in-plane compressibility, dln$a$/d$P$, 
manifesting the anisotropic nature of the crystal structure. Nonetheless, 
the bonding anisotropy in this material is not as large as other 
quasi-2D systems like alkali-metal intercalated graphite, and the 
elastic properties appear only moderately anisotropic. This information 
is valuable in testing the predictions of competing models for 
the mechanism of superconductivity. The diffraction experiments in 
combination with the data on the pressure dependence of $T_c$ permit 
us to determine its variation over a wide range of unit cell volume 
with an initial pressure coefficient, d(ln$T_c$)/d$V$= 0.18 $\AA^{-3}$.


The \mgb2 sample used in this work was prepared, as reported in \cite{nature},
by heating a pressed pellet of stoichiometric amounts of Mg and 
amorphous B for 10 hours at 700$^{\circ}$C under an argon pressure 
of 196 MPa and was superconducting with $T_c$= 39 K. 
Phase purity was confirmed by powder X-ray diffraction. The 
high pressure synchrotron X-ray diffraction experiments at ambient 
temperature were performed on beamline BL10XU at Spring-8, Japan. \mgb2 
was loaded in a diamond anvil cell (DAC), which was used for 
the high-pressure generation and was equipped with an inconel gasket. 
The diameter of the top face of the diamond culet was 1 mm, and the 
sample was introduced in a hole made in the gasket 0.2 mm deep and 0.4 
mm diameter. Silicone oil loaded in the DAC was used as a pressure medium. 
Pressure was increased at room temperature and was measured with 
the ruby fluorescence method. The diffraction patterns were collected 
using an image plate detector ($\lambda$= 0.49556 \AA) with 5 min 
exposure times. Integration of the two-dimensional diffraction images was 
performed with the local PIP software and data analysis with the Fullprof 
suite of Rietveld analysis programs \cite{fullprof}.


Synchrotron X-ray powder diffraction profiles of \mgb2 were 
collected at pressures between ambient and 6.15 GPa. Inspection 
of the diffraction data indicated that the pattern could be 
indexed as hexagonal at all pressures. Thus the same structural model 
(space group $P$6/$mmm$) was employed in the refinements of all datasets 
\cite{comment}. The Rietveld refinements (2$\theta$ range= 7-43$^{\circ}$
proceeded smoothly (Fig. 1), leading to values for the hexagonal 
lattice constants, $a$= 3.0906(2) \AA\ and $c$= 3.5287(3) \AA\ 
at ambient pressure (agreement factors: $R_{wp}$= 3.9\%, $R_B$= 5.9\%), 
and $a$= 3.0646(1) \AA\ and $c$= 3.4860(2) \AA\ at 6.15 GPa 
($R_{wp}$= 1.0\%, $R_B$= 7.0\%). Fig. 2(a) shows the pressure evolution 
of the volume of the unit cell of \mgb2 together with a least-squares 
fit of its ambient-temperature equation-of-state (EOS) to the semi-empirical 
second-order Murnagham EOS \cite{EOS}:

\begin{equation}
\label{1}
P=(K_0/K_0^{'})[(V_0/V)^{K_0^{'}}-1]
\end{equation}

where $K_0$ is the atmospheric pressure isothermal bulk modulus,
$K_0^{'}$ is its pressure derivative (=d$K_0$/d$P$), and $V_0$ is the unit
cell volume at zero pressure. The fit results in values of
$K_0$= 120(5) GPa and $K_0^{'}$= 36(3). The extracted value of the
volume compressibility, dln$V$/d$P$= 8.3(3)$\times$10$^{-3}$ GPa$^{-1}$
implies a stiff tightly-packed incompressible solid.

The anisotropy in bonding of the \mgb2 structure (Fig. 2 inset) 
is clearly evident in Fig. 2(b) which displays the variation of the hexagonal 
lattice constants $a$ and $c$ with pressure. As the applied pressure 
increases, the ($c$/$a$) ratio smoothly decreases (by $\sim$0.4\% to 
6.15 GPa). We described the pressure dependence of each lattice 
constant with a variant of eq. (1), in which $K_0$ and its pressure 
derivative, $K_0^{'}$ were substituted by the individual $K_x$ and 
$K_x^{'}$ ($x$= $a$, $c$) values. The results of these fits are 
also included in Fig. 2(b) and give: $K_a$= 410(20) GPa, $K_a^{'}$= 13(1); 
$K_c$= 292(12) GPa, $K_c^{'}$= 85(7). These values clearly reveal 
the diversity in bonding interactions present, with the solid 
being least compressible in the basal plane $ab$, in which the 
covalent B-B bonds lie (dln$a$/d$P$= 0.0024(1) GPa$^{-1}$). 
However, the interlayer linear compressibility, dln$c$/d$P$= 0.0034(1) 
GPa$^{-1}$ is only 1.4 times larger, implying very stiff Mg-B bonding; 
significantly, it is considerably smaller than those of the 
structurally related strongly anisotropic alkali-metal intercalated 
graphite ($cf.$ KC$_8$, dln$c$/d$P$= 0.0206 GPa$^{-1}$ \cite{graphite}.
Table 1 summarizes bond distance information at selected pressures.

The pressure coefficient of $T_c$ in \mgb2 is strongly negative, 
-dln$T_c$/d$P$$\sim$ 0.042 GPa$^{-1}$ \cite{highp}. It straddles the 
values of conventional non-cuprate $sp$- and $d$- superconductors 
($>$0.08 and $<$0.02 GPa$^{-1}$, respectively), while it is more 
than an order of magnitude smaller than those in fulleride superconductors 
(0.5 GPa$^{-1}$ for K$_3$\c60 \cite{k3c60} and 1.0 GPa$^{-1}$ 
for Na$_2$Cs\c60 \cite{mizuki}). Of course, the relevant comparison 
must be between the volume coefficients of $T_c$. Using the 
measured volume compressibility of \mgb2, we obtain d(ln$T_c$)/d$V$= 
0.18 $\AA^{-3}$. This is even larger than the "universal" value 
of alkali fullerides (0.07 $\AA^{-3}$ \cite{zhou}) and implies 
a very sensitive dependence of the superconducting properties to 
the interatomic distances. The results of converting the $V$($P$) data 
in Fig. 2(b) to $T_c$($V$) are shown in Fig. 3. Within the BCS 
formalism for superconductivity, $T_c$ $\propto$ 
($\hbar$$\omega_{ph}$)exp[-1/$JN$($\epsilon_F$)], where $\hbar$$\omega_{ph}$ 
is a phonon energy, $N$($\epsilon_F$) is the density-of-states at 
the Fermi level, and $J$ is the electron-phonon coupling constant. 
Given the highly incompressible nature of the \mgb2 solid, 
it is conceivable that the pressure coefficients of both $\hbar$$\omega_{ph}$
and $J$ are negligible. Then, to first order, Fig. 3 reflects 
the modulation of $T_c$ with decreasing $N$($\epsilon_F$), 
which results from the broadening of the conduction band as the 
B-B (and the Mg-B) orbital overlap increases with increasing external 
pressure. Such a situation is reminiscent of the \c60-based 
superconductors and suggests that higher $T_c$s may be also 
achieved, for the same band filling, by increasing the size 
of the unit cell volume by chemical substitution at the 
interstitial sites (negative pressure effect).

In conclusion, synchrotron X-ray powder diffraction at 
elevated pressures has been used to derive the evolution of the 
structural properties of the high-$T_c$ \mgb2 superconductor with 
pressure. In combination with the pressure dependence of $T_c$, 
these results provide the detailed dependence of the superconducting 
properties as both B-B and Mg-B bonding distances decrease and 
should form a stringent test of competing models for the 
interpretation of the superconducting pairing mechanism in this material. 
Detailed band structure calculations should be able to also shed 
light on the evolution of the charge transfer between 
Mg and B and decipher the relative importance between the 
boron layers and the interstitial metal ions to superconductivity.

KP thanks Monbusho for a Visiting Professorship to JAIST. We 
thank Spring-8 for the provision of synchrotron X-ray beamtime. 
The work at Aoyama-Gakuin University was partially supported by a 
Grant-in-Aid for Science Research from the Ministry of 
Education, Science, Sports and Culture, Japan and by a Grant 
from CREST. We acknowledge 
useful discussions with T. Arima, T. Takenobu and S. Suzuki.

\begin{table}[]
\small
\caption{Bond distances (\AA) for \mgb2 at selected pressures.}
\vspace{5mm}
\begin{tabular}{|l|l|l|l|}
Pressure (GPa)  & B-B &  Mg-B &  Mg-Mg 
\\
\\
ambient & 1.7844(1)      & 2.5094(1)   & 3.0906(2)   
\\
3.23  & 1.7738(1) & 2.4917(1) &  3.0724(1) 
\\
6.15 &  1.7694(1) &  2.4837(1) &  3.0646(1) 
\end{tabular}
\end{table}

\begin{figure}[tbp] \centering
\caption{Final observed (open circles) and calculated (solid line) 
ambient temperature synchrotron X-ray powder diffraction profiles 
of \mgb2 at 1.27 and 5.46 GPa in the range of 7$^{\circ}$ to 
43$^{\circ}$ ($\lambda$= 0.49556 \AA). The lower lines show the 
difference profiles and the vertical marks indicate the observed reflections.
} \label{fig1}
\end{figure}

\begin{figure}[tbp] \centering
\caption{Pressure dependence of: (a) the unit cell volume, $V$ and 
(b) the normalized hexagonal lattice constants ($a$/$a_0$) and 
($c$/$c_0$) of \mgb2 at ambient temperature. 
The lines through the data points are least-squares fits 
to the Murnaghan equation-of-state (eq. 1) in (a) and its 
variants in (b) (see text). Inset. Schematic diagram of the \mgb2 structure.
} \label{fig2}
\end{figure}

\begin{figure}[tbp] \centering
\caption{Superconducting transition temperature, $T_c$ versus 
unit cell volume, $V$ for \mgb2. The $V$($P$) data from Fig. 2(a) 
were converted to $T_c$($V$) by using the $T_c$($P$) data (solid 
circles) of ref. [8] to 1.84 GPa. The line through 
the points is a guide to the eye.
} \label{fig3}
\end{figure}


\begin{references}
\bibitem{comment}
The atoms were placed in the following positions in the unit cell: 
Mg in (1$a$) (0,0,0); B in (2$d$) (1/3,2/3,1/2).
\end{references}
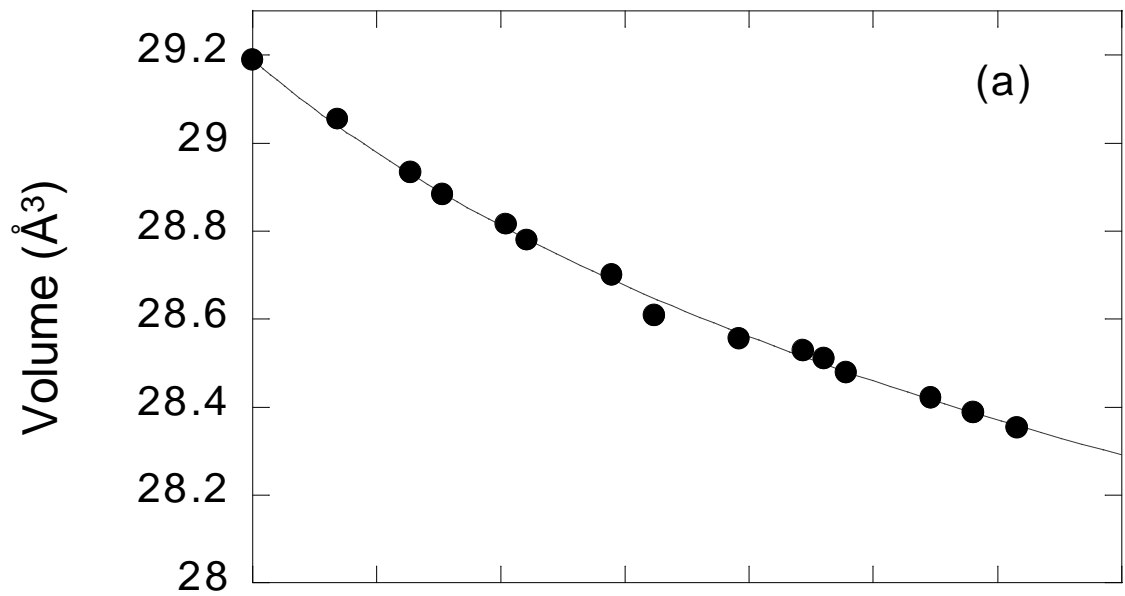
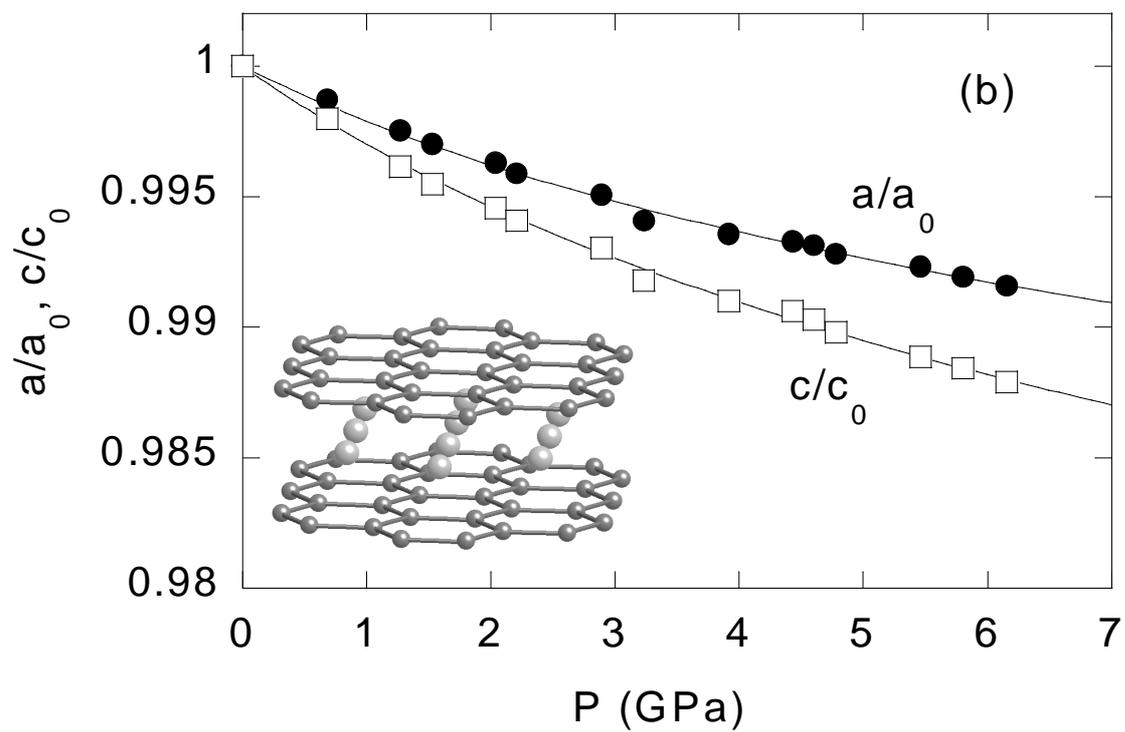
\end{document}